# Optimal Operation of Power Systems with Energy Storage under Uncertainty: A Scenario-based Method with Strategic Sampling

Ren Hu and Qifeng Li, *Senior Member, IEEE*

*Abstract*—The multi-period dynamics of energy storage (ES), intermittent renewable generation and uncontrollable power loads, make the optimization of power system operation (PSO) challenging. A multi-period optimal PSO under uncertainty is formulated using the chance-constrained optimization (CCO) modeling paradigm, where the constraints include the nonlinear energy storage and AC power flow models. Based on the emerging scenario optimization method which does not rely on pre-known probability distribution functions, this paper develops a novel solution method for this challenging CCO problem. The proposed method is computationally effective for mainly two reasons. First, the original AC power flow constraints are approximated by a set of learning-assisted quadratic convex inequalities based on a generalized least absolute shrinkage and selection operator. Second, considering the physical patterns of data and motived by learning-based sampling, the strategic sampling method is developed to significantly reduce the required number of scenarios through different sampling strategies. The simulation results on IEEE standard systems indicate that 1) the proposed strategic sampling significantly improves the computational efficiency of the scenario-based approach for solving the chance-constrained optimal PSO problem, 2) the data-driven convex approximation of power flow can be promising alternatives of nonlinear and nonconvex AC power flow.

## I. Introduction

Energy storage (ES) has been well-recognized for dealing with the challenges in power systems, such as shaving peak-load and filling valley-load. However, the current cost of battery ES is still expensive. According to the roadmap of ES issued by the U.S. department of Energy in 2020, by 2030 the levelized cost of battery ES may be reduced to only 10% of the current cost [1]. This probably makes ES widely used in power systems. However, the inter-temporal property of ES may couple the multi-period power system operation (PSO). Moreover, the intermittence of renewable energy (RE) brings the uncertainty to PSO. Hence, the exploration on optimizing the multi-period PSO with ES (PSO-ES) under the uncertainty

The authors are with the Department of Electrical and Computer Engineering, University of Central Florida, Orlando, FL 32816 USA (e-mail: hurenlaker@knights.ucf.edu, Qifeng.li@ucf,edu).

of RE is rather challenging. Unfortunately, the current deterministic approaches are incapable of capturing the uncertainty in the context of optimization. There exist some approaches of modeling optimization problems under uncertainty, such as stochastic, robust, and chance-constraint optimization methods [2]-[7]. The stochastic optimization [2], [3] attempts to find solutions with the best expected objective values based on the predefined probability distributions. The robust optimization [3], [4] enforces strict feasibility under the worst case, resulting in high conservativeness. Unlike the two methods above, the chance-constrained optimization (CCO) [5]-[7] can guarantee that the satisfactory probability of a solution is above a certain level if properly implemented. Generally, power system operators put higher weight in security than in cost-saving. As a tradeoff, power operators may be more interested in solutions with the low probability of constraint violation. Therefore, in this paper, CCO is adopted to model the multi-period PSO-ES problem under uncertainty (CC-PSO-ES). Despite its widespread applications in engineering disciplines, the original CCO is generally computationally expensive for large-scale systems like power grids. Additionally, the conventional solution methods of CCO overly depend on actual joint probability distribution function (PDF) of random variables which is hard to access [5]-[7].

As an alternative to the PDF-based methods to solve the CCO problems, the scenario-based solution method, called scenario optimization has been applied in probabilistic optimization problems [8], [9], learning models and artificial intelligence (AI) [10], [11]. The key to scenario optimization aims at how to determine the minimum sample size (MSS) required to satisfy the specific probability level [8], [12], [13]. Reference [8] illustrates a random sampling (RS)-based method (RSM) to estimate the MSS associated with the number of decision variables under the convex program. However, for complex systems with numerous decision variables, the MSS estimated by RSM may explode as the MSS is proportional to the size of decision variables [8]. To tackle this issue, the fast algorithm for scenario technique (FAST), a two-stage method [10] has been proposed to cut down the sample size and applied in computing optimal power flow with uncertainty [14]. As stated in RS-based methods [9],[10], there may have a small size of 'active scenarios' that essentially decides the optimal solution. The number of active scenarios is proven to be at most the decision variable size, which is far smaller than the sample size determined by RS-based methods. In other words, most of scenarios selected by RS-based methods are 'inactive' and removable. However, these active scenarios are unknown before solving the RS-based optimization problems. Inspired by the resounding sequential sampling [10], [11] used in machine learning and the existence of physical patterns in data, a concept of strategic sampling is developed in this paper to find a much smaller size of scenarios that can approximate the effect of the active scenarios before optimization, through physics-guided sampling [32], [33], dissimilarity-based learning [15] and reinforcement learning [16] methods.

The above-mentioned scenario-related optimization methods are currently only applicable to convex program problems [9], [10]. However, the constraints of AC power flow (ACPF) in CC-PSO-ES problem are inherently nonlinear and nonconvex [6]. Currently, the approximations of ACPF have been discussed from the perspectives of linearization and convexification. The linear approximations like the DC model [7], [17] and other linear ACPF [6], [18], are generally easy to solve, however, many of which ignore the quadratic terms of voltages resulting in inaccuracy of model. The typical convex approximations have been widely studied, such as the second-order cone (SOC) [19], semi-definite programming (SDP) [20], convex DistFlow [21], quadratic convex (QC) [22], moment-based [23], convex hull relaxation [24], and the learning-based convex approximation [25], [26]. References [26]-[28] found that the SDP relaxation may not guarantee the exactness of solutions and its exactness greatly depends on the critical assumptions of network topologies and physical parameter settings. In [26], the authors developed an ensemble learning-based data-driven convex quadratic approximation (DDCQA) of ACPF with higher accuracy and efficiency than the SDP relaxation. This paper introduces the generalized least absolute shrinkage and selection operator (LASSO) [29], [30] to improve the DDCQA developed in [26] from both aspects of computing time and space used.

To solve the CC-PSO-ES which is a complex multi-period nonlinear nonconvex optimization problem, this paper proposes a novel scenario-based solution method based on the DDCQA of ACPF and strategic sampling. The proposed approach is more computationally effective using only few effective scenarios, compared with RSM. The contributions of this paper are written as below:

1) The strategic sampling is developed based on physics-guided sampling and learning-based sampling methods to select a small size of scenarios for solving CC-PSO-ES problem.

2) The DDCQA of ACPF is improved by generalized LASSO from the aspects of computational time and space complexity, then applied to convert the originally intractable nonconvex CC-PSO-ES problem into a tractable convex quadratic optimization problem.

The rest of this paper is organized as follows: Section II illustrates the formulations of deterministic and chance-constrained multi-period PSO-ES problems. In Section III, scenario optimization is introduced, and the novel scenario-based solution method for CC-PSO-ES problem is developed through strategic sampling and the DDCQA of ACPF modified by generalized LASSO. The empirical IEEE case analyses and conclusions are displayed in Section IV and V, respectively.

## II. PROBLEM FORMULATION

This section formulates the optimal multi-period operation for power systems with energy storage under the modeling paradigm of chance-constrained optimization step-by-step.

### A. Deterministic Multi-period PSO with Battery Energy Storage

In an $n$-bus power system, the deterministic formulation of the multi-period PSO-ES is given as following which can also be considered as a multi-period AC optimal power flow (ACOPF) with adjustable generation and battery energy storage:

$$Min \sum_{t \in T} \sum_i (c_{i1} P_{i,t}^G + c_{i2} (P_{i,t}^G)^2) \tag{1a}$$

$$s.t. \quad e_{i,t} \sum_{j=1}^n (G_{ij} e_{j,t} - B_{ij} f_{j,t}) + f_{i,t} \sum_{j=1}^n (G_{ij} f_{j,t} + B_{ij} e_{j,t}) = P_{i,t}^G - P_{i,t}^{net} + P_{ES,i,t} \tag{1b}$$

$$f_{i,t} \sum_{j=1}^n (G_{ij} e_{j,t} - B_{ij} f_{j,t}) - e_i \sum_{j=1}^n (G_{ij} f_{j,t} + B_{ij} e_{j,t}) = Q_{i,t}^G - Q_{i,t}^{net} + Q_{ES,i,t} \tag{1c}$$

$$G_{ij} e_{i,t} e_{j,t} - B_{ij} e_{i,t} f_{j,t} + B_{ij} f_{i,t} e_{j,t} + G_{ij} f_{i,t} f_{j,t} - G_{ij}(e_{i,t}^2 + f_{i,t}^2) = P_{ij,t} \tag{1d}$$

$$G_{ij} f_{i,t} e_{j,t} - B_{ij} e_{i,t} e_{j,t} - G_{ij} e_{i,t} f_{j,t} - B_{ij} f_{i,t} f_{j,t} + B_{ij}(e_{i,t}^2 + f_{i,t}^2) = Q_{ij,t} \tag{1e}$$

$$P_{ES,i,t}^{loss} V_{i,t}^2 = r_i^{eq} P_{ES,i,t}^2 + r_i^{cvt} Q_{ES,i,t}^2 \tag{1f}$$

$$P_{ES,i,t}^{net} = P_{ES,i,t} + P_{ES,i,t}^{loss} \tag{1g}$$

$$P_{ES,i,t}^2 + Q_{ES,i,t}^2 = S_{ES,i}^{max2} \tag{1h}$$

$$E_{ES,i}^{min} \le E_{ES,i,0} + \Delta t \sum_{k=1}^t P_{ES,i,k}^{net} \le E_{ES,i}^{max} \tag{1i}$$

$$V_{i,t}^{min2} \le e_{i,t}^2 + f_{i,t}^2 \le V_{i,t}^{max2} \tag{1j}$$

$$P_{i,t}^{Gmin} \le P_{i,t}^G \le P_{i,t}^{Gmax} \tag{1k}$$

$$Q_{i,t}^{Gmin} \le Q_{i,t}^G \le Q_{i,t}^{Gmax} \tag{1l}$$

$$P_{ij,t}^2 + Q_{ij,t}^2 \le S_{ij,t}^{Gmax} \tag{1m}$$

where $i, j = 1, 2, \ldots, n$; $T$ is the index set of time periods; the subscript $t$ denotes the $t$-th hour; $c_{i1}$, $c_{i2}$ are the generator cost coefficients; $P_{i,t}^G$, $Q_{i,t}^G$ are the generator active and reactive power; $P_{i,t}^{net}$, $Q_{i,t}^{net}$ are the net active and reactive power inputs of power loads and renewable energy outputs; $P_{i,t}^{Gmin}$, $P_{i,t}^{Gmax}$, $Q_{i,t}^{Gmin}$, $Q_{i,t}^{Gmax}$ are the lower and upper limits of the generator active and reactive power; $V_{i,t}^{min}$, $V_{i,t}^{max}$ are the lower and upper limits of bus voltage; $S_{ij,t}^{Gmax}$ is the branch transmission capacity; $e_{i,t}$, $f_{i,t}$ represent the real and imaginary parts of voltage; $P_{ij,t}$, $Q_{ij,t}$ denote the active and reactive line flow; $G_{ij}$, $B_{ij}$ are the real and imaginary parts of the line admittance; $P_{ES,i,t}$, $Q_{ES,i,t}$ are the active and reactive power of energy storage; $P_{ES,i,t}^{loss}$ is the active power loss of energy storage; $r_i^{eq} = r_i^{batt} + r_i^{cvt}$, $r_i^{cvt}$ are the equivalent resistances of the battery and converter [24]; $V_{i,t}^2$ is the squared magnitude of voltage that approximates to 1.0 pu; $P_{ES,i,t}^{net}$ is the net active power of energy storage; $E_{ES,i}^{min}$, $E_{ES,i}^{max}$ are the capacity limits of energy storage; $E_{ES,i,0}$ is the initial energy status of energy storage; $S_{ES,i}^{max}$ is the maximum apparent power of energy storage.

### B. Chance-Constrained Formulation of Multi-period PSO with Adjustable Generation and Battery Energy Storage

Considering that the net load inputs of power loads and renewable energy generations (PLRES) are random variables, the power generation may be fluctuating, which consists of the base and adjustable parts. The base part meets the forecast net demand injection of PLRES. The gap between the forecast and actual net demand injections is satisfied by the adjustable part. According to the affine control scheme [31], then

$$P_{j,t}^{net,s} = P_{j,t(frt)}^{net} + \Delta p_j^s \tag{2a}$$
$$Q_{j,t}^{net,s} = Q_{j,t(frt)}^{net} + \Delta q_j^s \tag{2b}$$
$$P_{i,t}^{G,s} = P_{i,t(base)}^G - \omega_{pi} \sum_{j=1}^n \Delta p_j^s \tag{2c}$$
$$Q_{i,t}^{G,s} = Q_{i,t(base)}^G - \omega_{qi} \sum_{j=1}^n \Delta q_j^s \tag{2d}$$
$$P_{ES,i,t}^s = P_{ES,i,t(base)} - \omega_{pi} \sum_{j=1}^n \Delta p_j^s \tag{2e}$$
$$Q_{ES,i,t}^s = Q_{ES,i,t(base)} - \omega_{qi} \sum_{j=1}^n \Delta q_j^s \tag{2f}$$
$$\sum_i \omega_{pi} + \sum_j \omega_{pj} = 1 \tag{2g}$$
$$\sum_i \omega_{qi} + \sum_j \omega_{qj} = 1 \tag{2h}$$

Where $S$ is the scenario sets, $s \in S$; $P_{i,t(base)}^G$, $Q_{i,t(base)}^G$ are the base parts of the generator's active and reactive power; $P_{ES,i,t(base)}$, $Q_{ES,i,t(base)}$ are the base parts of the energy storage's active and reactive power; $\omega_{pi}$, $\omega_{qi}$ are the participation factors of the generator's or energy storage's active and reactive power, respectively; $P_{j,t(frt)}^{net}$, $Q_{j,t(frt)}^{net}$ denote the forecast values of the net active and reactive load inputs of PLRES, respectively; $\Delta p_j^s$, $\Delta q_j^s$ denote the corresponding forecast errors of $P_{j,t}^{net,s}$ and $Q_{j,t}^{net,s}$, respectively. Note that we assume the difference between the base case and real-time power loss can be compensated by the reference generator and it is typically negligible. Suppose that the chance-constraint method is used to model the problem (1) under uncertainty. Then, by substituting (2a)~(2h) into (1) and updating each variable in (1) with the superscript $s$, the deterministic problem (1) is reformulated into a CC-PSO-ES problem in (2):

$$\min_{s \in S} E[\sum_{t \in T} \sum_i (c_{i1} P_{i,t}^{G,s} + c_{i2}(P_{i,t}^{G,s})^2)] \tag{2i}$$
$$s.t. \ Pr(h(y,\delta) = 0) \geq 1 - \alpha \tag{2j}$$
$$Pr(f(y,\delta) \leq 0) \geq 1 - \alpha \tag{2k}$$

where $h(y,\delta)$, $f(y,\delta)$ compacts the constraints of (1b)~(1h) and (1i)~(1m), respectively; $y$ is the variable vector consisting of the decision variables $P_{i,t}^{G,s}$, $Q_{i,t}^{G,s}$, and the state variables such as the bus voltage; $\delta$ is the random variable vector such as the power loads and renewable energy generations, $P_{j,t}^{net,s}$ and $Q_{j,t}^{net,s}$; $Pr(\cdot)$ enforces each constraint at a specific confidence level; $\alpha$ is the probability level.

## III. A SCENARIO-BASED SOLUTION METHOD WITH STRATEGIC SAMPLING AND DATA DRIVEN CONVEX APPROXIMATION

### A. Scenario Optimization

Scenario optimization has been widely used in machine learning [10], [11], whose general idea is to use a finite number of scenarios to approximate the probabilistic constraints (2f) and (2g) with a specific confidence level. The mathematical formulation can be represented as

$$Min \ C^T v \tag{3a}$$
$$s.t. \ F(v, \delta^{(s)}) \leq 0, (s = 1,2,..,N') \tag{3b}$$

where (3a) is a linear objective function related to the decision variable vector $v$; $\delta^{(s)}$ denotes the $s$-th scenario sampled from the uncertainty set; $F$ is a convex function on $v$; $N'$ is the estimated number of scenarios. In the existing applications of scenario optimization, the constraints depend upon the random samples, and the sample size discussed in the statistical learning [12], [13] has a conversative estimate unrelated to the number of decision variables. Until in reference [8], the lower bound of the sample size related to the decision variable size under convex program settings is derived from the aspect of binomial distribution. A theorem in [8] states that, if $N'$ is sufficiently large, the optimal solution of (3) can satisfy the chance constraints (2j)~(2k). Scenario optimization is still an emerging solution method for chance-constrained optimization that does not rely on pre-known PDFs [8]-[11]. The RS-based method in [8] provides some discussions on how to determine the required number of scenarios, shown in (3c):

$$N' \geq \frac{2}{\epsilon}\left(ln\frac{1}{\beta} + d'\right) \tag{3c}$$

where $d'$ is the dimension of decision variable vector; $\epsilon \in (0,1)$, $1 - \beta \in (0,1)$ are the violation probability level and confidence level, respectively. However, in CC-PSO-ES problem, the number of scenarios required by the RS-based method may be large, which results in significant challenge in computation. Moreover, scenario optimization is designed for convex optimization, while there have nonconvex constraints in current CC-PSO-ES problem. Hence, the following sections will discuss how to tackle the two issues through the strategic sampling and DDCQA, respectively.

### B. Strategic Sampling

Instead of random sampling with plentiful inactive scenarios [8], we attempt to develop a framework of strategic sampling to find out a smaller number of effective scenarios that include the active ones. According to different selection strategies, there may have diverse specific strategic sampling methods shown in Fig.1, such as physics-guided sampling (PGS), learning-based sampling, and hybrid sampling, etc. [15], [16], [32], [33]. The PGS is designed considering there might have specific patterns in PLRES data. The patterns may be related to the temporal, spatial, and meteorological conditions [32], [33]. The learning-based sampling is based on machine learning methods, such as dissimilarity-based learning and reinforcement learning (RL) [15], [16]. The hybrid one may be the combination of any two sampling methods. In this research, two types of two-stage hybrid sampling methods are developed and named as HS1 and HS2. The first stage is the physics-guided sampling (PGS). Then, at the second stage, i.e., the stage of learning-based sampling, one of dissimilarity-based sampling (DBS) and RL-based sampling (RLS) will be chosen to select the $d$ dissimilar samples. As a rule of thumb, in IEEE-5 system, $d$ is suggested to be the number of the decision variables; all other systems can set $d$ to be 10% of the decision variable size.

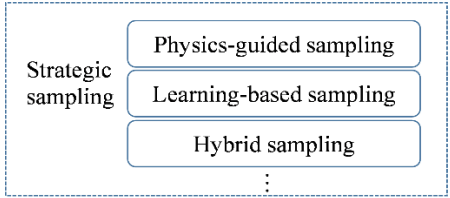

Fig.1 Framework of Strategic Sampling

1. Physics-guided Sampling

Assume there are regional power systems like IEEE-5, -9, -57, -118 systems. The gist of PGS for a specific regional power system can be described in Fig.2 [32], [33].

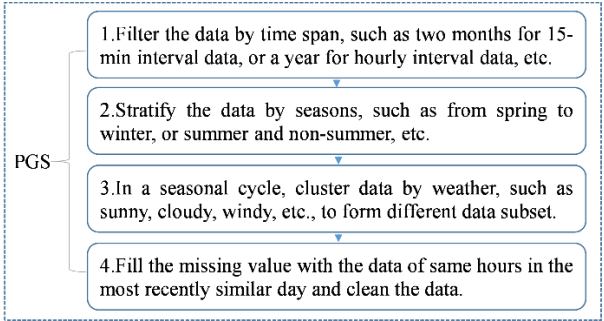

Fig.2 The Flowchart of Physics-guided Sampling

2. Dissimilarity-based Sampling

The dissimilarity-based sampling (DBS) is defined to use a learning function to select the scenarios. The purpose of using a learning function is to ensure that the scenarios selected should be dissimilar enough to maximize the scenario difference in a specific data subset obtained after PGS. To measure the dissimilarity of samples, in machine learning the distance metrics are commonly used, such as the Euclidean distance [10], [11], [15]. Assume that $\delta_i, \delta_j$ are the $i$-th and $j$-th samples where $\delta_i = (\delta_i^1, \delta_i^2, \ldots, \delta_i^m)$; $m$ is the dimension of each sample. The dissimilarity between two samples can be measured by the Euclidean distance calculated as

$$D_{ij} = \sqrt{\sum_{k=1}^{m}(\delta_i^k - \delta_j^k)^2} \tag{4a}$$

where $D_{ij}$ denotes the dissimilarity of two samples, used for determining the new samples. The specific hybrid sampling that consists of PGS and DBS is described as HS1 in Fig.3 where $N$ is computed by FAST.

3. Reinforcement Learning-based Sampling

The RL-based sampling (RLS) [16] considers each sample as a state $s_i$, selecting each sample as an action $a_i$, and the probability of each transition from the $i$-th state $s_i$ to $j$-th state $s_j$ as $\pi_{ij}$ defined as:

$$\pi_{ij}(s_j|s_i, a_i) = \frac{D_{ij}}{\sum_{j \in H} D_{ij}} \tag{4b}$$

where the indices $i$ and $j$ correspond to the current and next state; $H$ is the potential state set for next state. The reward of each transition is defined to be proportional to $D_{ij}$, and the discount factor of reward is set to one in this research. The implementation of RLS is shown as HS2 in Fig.4. The main difference between DBS and RLS lies in that RLS only consider the dissimilarity between the current selected sample and the candidate sample, while DBS computes the average dissimilarity between the previous all selected samples and the candidate sample.

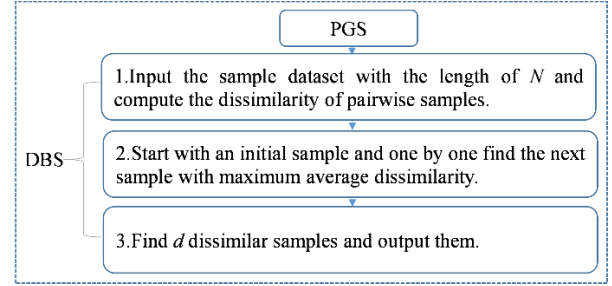

Fig.3 The Flowchart of Hybrid Sampling Type 1 (HS1)

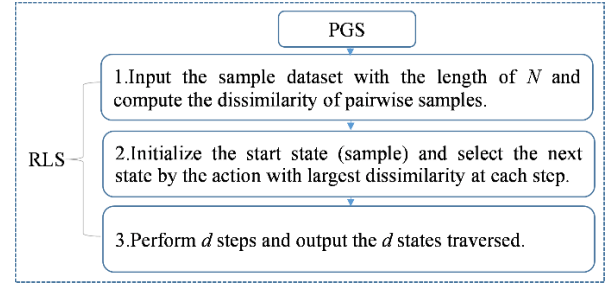

Fig.4 The Flowchart of Hybrid Sampling Type 2 (HS2)

C. Data-Driven Convex Quadratic Approximation of Power Flow

Note that the scenario-based solution methods discussed so far are designed for the convex program [8], [9], while, in the original CC-PSO-ES problem, the constraints of ACPF are nonconvex. To deal with the issue of nonconvexity, we improve the data-driven convex quadratic approximation (DDCQA) of ACPF [26] using the generalized LASSO [29], [30]. For the ease of analysis, the constraints (1b)-(1e) at $t$-th hour are reformulated as the following matrix form:

$$P_{i,t} = X_t^T A_i X_t \tag{5a}$$
$$Q_{i,t} = X_t^T B_i X_t \tag{5b}$$
$$P_{ij,t} = X_{ij,t}^T A_{ij} X_{ij,t} \tag{5c}$$
$$Q_{ij,t} = X_{ij,t}^T B_{ij} X_{ij,t} \tag{5d}$$
$$X_t = [x_{1,t}\ x_{2,t}, \ldots, x_{2n,t}]^T = [e_{1,t}\ f_{1,t}, \ldots, e_{n,t}\ f_{n,t}]^T \tag{5e}$$
$$X_{ij,t} = [x_{2i-1,t}\ x_{2i,t}\ x_{2j-1,t}\ x_{2j,t}]^T = [e_{i,t}\ f_{i,t}\ e_{j,t}\ f_{j,t}]^T \tag{5f}$$

Where $A_i, B_i, A_{ij}, B_{ij}$ are symmetric indefinite matrices consisting of elements of the admittance matrix, implying all de-pendent variables $P_{i,t}$, $Q_{i,t}$, $P_{ij,t}$, and $Q_{ij,t}$ are nonconvex functions of the independent variables $X_t$ or $X_{ij,t}$. Hence, a convex quadratic mapping (5g)-(5j) between power, i.e., $P_{i,t}$, $Q_{i,t}$, $P_{ij,t}$, and $Q_{ij,t}$, and voltage $X_t$ or $X_{ij,t}$, is defined as:

$$P_{i,t} = X_t^T A_i^p X_t + B_i^p X_t + c_i^p \tag{5g}$$
$$Q_{i,t} = X_t^T A_i^q X_t + B_i^q X_t + c_i^q \tag{5h}$$
$$P_{ij,t} = X_{ij,t}^T A_{ij}^p X_{ij,t} + B_{ij}^p X_{ij,t} + c_{ij}^p \tag{5i}$$
$$Q_{ij,t} = X_{ij,t}^T A_{ij}^q X_{ij,t} + B_{ij}^q X_{ij,t} + c_{ij}^q \tag{5j}$$

Where $A_i^*$, $A_{ij}^*$ denote positive semi-definite (PSD) coefficient matrices of the quadratic terms, respectively; $B_i^*$, $B_{ij}^*$ denote coefficient vectors of the linear terms; $c_i^*$, $c_{ij}^*$ denote constant terms; the upper index (*) includes the set $\{p, q\}$ indicating the active or reactive power.

According to [26], the PSD matrices in (5g)-(5j) can be obtained via training historical data using the polynomial regression as a basic learner to learn the convex relationships between the voltage and the active or reactive power. Then, ensemble learning methods are used to assemble all basic learners, to boost the performance of model. However, the PSD matrices in (5g)-(5j) are dense and high-dimensional, which is an obstacle in computing the complex CC-PSO-ES problem. Therefore, the generalized LASSO [29], [30] is introduced to learn more compact and sparser PSD matrices for the purposes of speeding up the computational efficiency and saving the storage space. The following illustration of generalized LASSO takes $P_{i,t}$ as an example based on dataset $\{x_{it}, y_{it}\}_{i=1}^M$ where $x_{it}, y_{it}$ are the $i$-th observed voltage input and active power output at $t$-th hour; $M$ is the training sample size. The detailed formulation is written as

$$min \frac{1}{M}\sum_{i=1}^M (y_{it} - X_t^T A_i^p X_t - B_i^p X_t - c_i^p)^2 + \mu \sum_{j=1}^{M'}|\theta_j| \tag{5k}$$
$$s.t. \; A_i^p \succcurlyeq 0 \tag{5l}$$

where $\theta_j$ is the $j$-th coefficient constituted by the entries of $A_i^p$; '$\succcurlyeq$' means $A_i^p$ is a PSD matrix; $M'$ is the number of coefficients; $\mu \geq 0$ is a tunable regularization parameter that controls the degree of shrinkage. By the shrinkage, some coefficients may be zero, which means the matrix $A_i^p$ becomes sparse.

### D. Convex Hull Relaxation of Energy Storage Model

As the magnitude of voltage has little change, we assume that $V_{i,t}^2 \approx 1.0$ pu in (1f). Then, the convex hull relaxation [24] of the energy storage model can be formulated as the followings:

$$\|Jz_{i,t}\|_2 - b^T z_{i,t} \leq 0 \tag{6a}$$
$$\|J_i z_{i,t}\|_2 - b^T z_{i,t} \leq m_i \tag{6b}$$
$$k_i^T z_{i,t} - 2m_i \leq 0 \tag{6c}$$

where $J = diag([\sqrt{2} \; \sqrt{2} \; 1 \; 1]^T)$, $b = [0 \; 0 \; 1 \; 1]^T$, $J_i = diag([0 \; \sqrt{2r_i^{batt}} \; 1 \; 1]^T)$, $z_{i,t} = [P_{ES,i,t} \; Q_{ES,i,t} \; P_{ES,i,t}^{loss} \; 1]^T$,

$k_i = [0 \; 0 \; r_i^{eq} \; S_{ES,i}^{max2}]^T$, $m_i = r_i^{eq} S_{ES,i}^{max2}$.

### E. Scenario Programming Formulation of CC-PSO-ES

Based on the DDCQA of power flow constraints in (5g)-(5j) and the convex hull relaxation of energy storage model in (6a)-(6c), the corresponding convex constraints are written as

$$X_t^T A_i^p X_t + B_i^p X_t + c_i^p \leq P_{i,t}^G - P_{i,t}^{net} + P_{ES,i,t} \tag{7a}$$
$$X_t^T A_i^q X_t + B_i^q X_t + c_i^q \leq Q_{i,t}^G - Q_{i,t}^{net} + Q_{ES,i,t} \tag{7b}$$
$$X_{ij,t}^T A_{ij}^p X_{ij,t} + B_{ij}^p X_{ij,t} + c_{ij}^p \leq P_{ij,t} \tag{7c}$$
$$X_{ij,t}^T A_{ij}^q X_{ij,t} + B_{ij}^q X_{ij,t} + c_{ij}^q \leq Q_{ij,t} \tag{7d}$$
$$x_{2i-1,t}^2 + x_{2i,t}^2 \leq V_{i,t}^{max2} \tag{7e}$$

(1k)-(1m) and (6a)-(6c).

Then, we introduce an auxiliary variable $z$ used to reformulate the objective function (2i) into a linear formulation (7f) with a convex constraint (7g) as

Minimize: $z$ (7f)

$$s.t. \; \sum_{s \in S}\sum_{t \in T}\sum_i (c_{i0} + c_{i1}P_{i,t}^{G,s} + c_{i2}(P_{i,t}^{G,s})^2) \leq |S|z \tag{7g}$$

where $|S|$ is the number of scenarios considered. As the constraints above are all convex, the scenario programming, i.e., the strategic sampling-based solution approach for the CC-PSO-ES problem can be rewritten in (8) as

Minimize: (7f) (8a)

$$s.t. \; F'(y, \delta^{(s)}) \leq 0, (s = 1,2,\dots,d) \tag{8b}$$

where $F'(y, \delta)$ compacts all constraints in (7a)-(7e), (7g), (1k)-(1m), and (6a)-(6c) at each scenario. The corresponding sampling procedure for the problem (8) is illustrated in the section of strategic sampling.

## IV. SIMULATION ANALYSIS

### A. Case selection and Data Collection

The real-world power systems and their data are expected to use in this research. However, they are not available in public. As empirical alternatives, some IEEE standard test systems such as IEEE-5, -9, -57, -118 systems and the relevant simulating data are applied. The net active and reactive power loads at each bus are based on the hourly load curves of ISO new England and set up at the range of [0.7, 1.3] of their true values to simulate the uncertainty of power loads and renewable energy generations and to generate the 24-hour simulating data. The settings of energy storage units are summarized in TABLE I. Considering sampling in the whole sample space may be computationally expensive, for each test system, the sample size determined by the RS-based method [8] is treated as the sample space. The goal is to demonstrate the efficacy of the proposed solution method with fewer effective scenarios via the DDCQA of ACPF and the strategic sampling methods. The simulations are performed in Matlab with cvx package.

TABLE I
The Settings of Energy Storage Units

| Case | IEEE-5 | IEEE-9 | IEEE-57 | IEEE-118 |
|---|---|---|---|---|
| Units | 2 | 2 | 3 | 3 |

| Bus No. | 3, 5 | 5, 7 | 8, 9, 12 | 59, 90, 116 |
|---|---|---|---|---|
| Capacity | 1MVA, 2MWh | 0.75MVA, 1.5MWh | 0.75MVA, 1.5MWh | 1MVA, 2MWh |

### B. Computational Complexity Comparison of DDCQA

To compare the computational efficiency of the DDCQA of ACPF before and after improvement, we explore the average training time for the active and reactive power at each bus and the average computation time for solving the CC-PSO-ES problem on test systems, using the method in [26] named as 'old' and the one improved by the generalized LASSO named as 'new', shown in Figs.5 and 6. Figs.5 and 6 indicate that there exist significant improvements in both the training time and computing time of CC-PSO-ES problem, before and after using generalized LASSO. Particularly, on IEEE-57 system, it only takes about 25% of the original average training time to train the improved DDCQA and about 40% of the original average computing time of CC-PSO-ES to obtain the solution. For the IEEE-118 system, the average training time and the average computing time of CC-PSO-ES used now are only about 2% and 5% of the ones before, respectively. Moreover, the storage space usage of the matrix $A_i^p$ before and after improving DDCQA on all test systems has been displayed in Fig.7. Similarly, Fig.7 shows, that on all test systems, the storage space consumed now is reduced by over 75% compared with the one consumed before. For IEEE-57 and -118 systems, the storage space used now may only account for 1%-2% of the one used before. Overall, the improved DDCQA based on the generalized LASSO greatly degrades the computational complexity in training and computing the optimization problem.

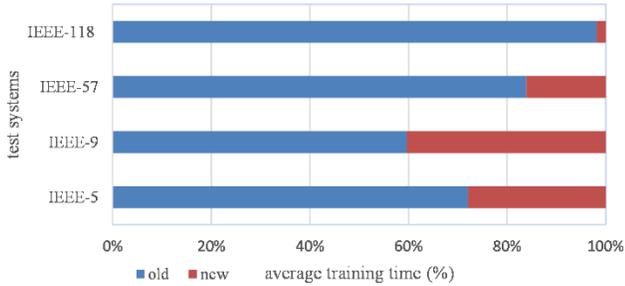
Fig.5 Comparison of Training Time

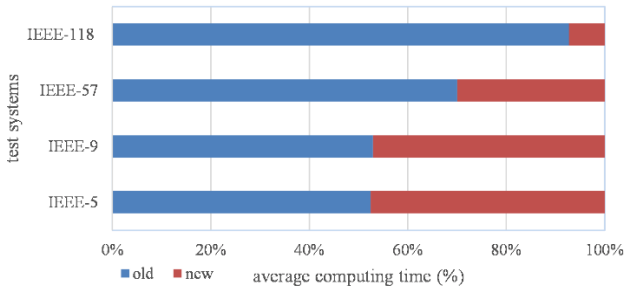
Fig.6 Comparison of Computing Time of CC-PSO-ES

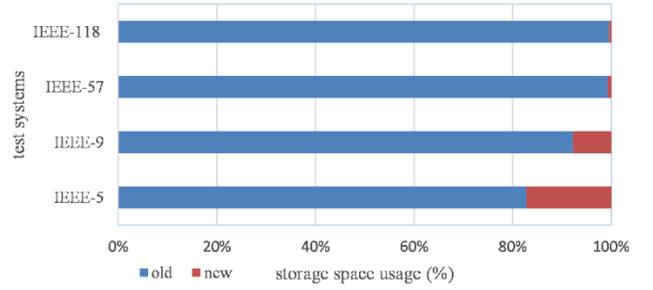
Fig.7 Storage Space Usage Comparison of Matrix

### C. Performance Comparison of Solution Methods

In this numerical experiment, we set the allowed violation probability level and the confidence level to be $\epsilon = 0.05$ and $1 - \beta = 0.9999$, respectively. The estimated number of scenarios required by the RS-based method [8] and FAST, and the decision variable size can be shown in TABLE II as below. Meanwhile, the total costs without and with considering the uncertainty of renewable energy and power load (REPL) are computed respectively, as 'Base Cost' (BC) and 'Objective Cost' (OC) in TABLE III. The 'Objective Cost' is computed through FAST as the benchmark.

TABLE II
Sample Sizes of FAST and RSM and the Size of Decision Variable

| Case | IEEE-5 | IEEE-9 | IEEE-57 | IEEE-118 |
|---|---|---|---|---|
| $d'$ | 864 | 1104 | 5904 | 17328 |
| $N$ | 1050 | 1290 | 6090 | 17514 |
| $N'$ | 34929 | 44529 | 236529 | 693489 |
| Ratio1= $d'/N'$ | 0.02474 | 0.02479 | 0.02496 | 0.02499 |

TABLE III
Total Costs without and with the Uncertainty of REPL

| Case | IEEE-5 | IEEE-9 | IEEE-57 | IEEE-118 |
|---|---|---|---|---|
| Base Cost ($/h) | 451710 | 229074 | 280175 | 3509250 |
| Objective Cost ($/h) | 477487 | 262187 | 296260 | 3603450 |
| Ratio2=OC/BC | 1.0570 | 1.1445 | 1.0574 | 1.0268 |

TABLE II indicates that the number of active scenarios only accounts for at most 2.5% of the sample size computed by the RS-based method. In other words, the majority of $N'$ samples may be useless for solving the CC-PSO-ES problem. Compared with RS-based method, FAST greatly reduces the number of scenarios to from $N'$ to $N$. Especially, for IEEE-57 and -118 systems, the sample sizes in multi-period situation required by RS-based method are more than 230k and 690k, respectively. It will be difficult to solve the CC-PSO-ES problems with such large number of scenarios in practice. The ratio by the objective cost and base cost in TABLE III demonstrates that in this research the (±30%) uncertainty of REPL may increase the total cost by 2%~15%.

To solve the CC-PSO-ES problems, the strategic sampling methods, i.e., two-stage hybrid sampling methods proposed are used. In practice, the first stage of physics-guided sampling (PGS) should be implemented at first. Then, the second stage of learning-based sampling methods, i.e., DBS and RLS, will be applied directly to determine the effective sample size. The difference between two hybrid sampling methods resides in the stage of learning-based sampling. As the initial sample can affect the sample size selected by RLS and DBS, the experimental simulation starting with different initial samples is explored. The best and worst sample selections by RLS and DBS, i.e., their minimum and maximum sample sizes for solving the CC-PSO-ES problems, are computed, and shown in TABLE IV. From TABLE IV, we can infer that:

1) The hybrid sampling methods through RLS and DBS can further reduce the effective sample size, i.e., the sample size required by the scenario-based methods can be far less than the ones with the RS-based method and FAST. For instance, in IEEE-5 system the solution methods through RLS and DBS find the optimal solution within 890 and 450 scenarios, respectively, which is smaller than 1050 scenarios determined by FAST. In a similar fashion, in IEEE-9, -57 and -118 systems, 120, 800 and 1800 scenarios are large enough for the solution methods through RLS and DBS to reach the optimal solution.

2) Under both best and worst sample selections, DBS outperforms RLS almost on four test systems. More specifically, the solution method through DBS finds the optimal solution of CC-PSO-ES problem more efficiently with less scenarios than the one through RLS, except that for the best cases in IEEE-9 and -118 systems DBS and RLS perform equally well. The main reason may be that DBS selects each sample based on the maximum average dissimilarity between the candidate sample and the previous all selected ones, while RLS selects samples through the maximum dissimilarity between the candidate sample and the most recently selected one.

TABLE IV
The Best and Worst Sample Selections by RLS and DBS

| Case | IEEE-5 | IEEE-9 | IEEE-57 | IEEE-118 |
|---|---|---|---|---|
| RLS(best) | 223 | 2 | 4 | 2 |
| RLS(worst) | 883 | 114 | 729 | 1780 |
| DBS(best) | 3 | 2 | 3 | 2 |
| DBS(worst) | 432 | 96 | 289 | 637 |

*D. Verification of Learning-based Sampling Methods*

To verify the DBS and RLS, at each round a new scenario is added sequentially until the number of scenarios reaches to $N$. In the actual application of DBS and RLS, there is no need to repeat the verification process. The corresponding results of CC-PSO-ES problems on four IEEE test cases are computed and compared based on DBS and RLS, displayed in Fig. 8~11.

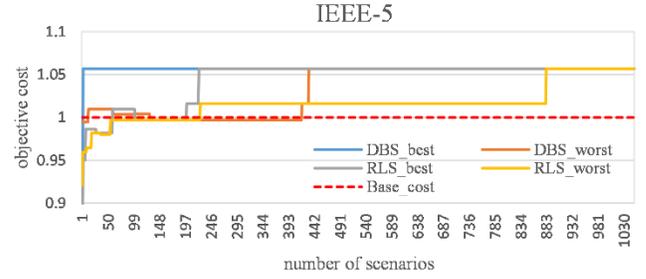

Fig.8 Performance Comparison of RLS and DBS in IEEE-5 System

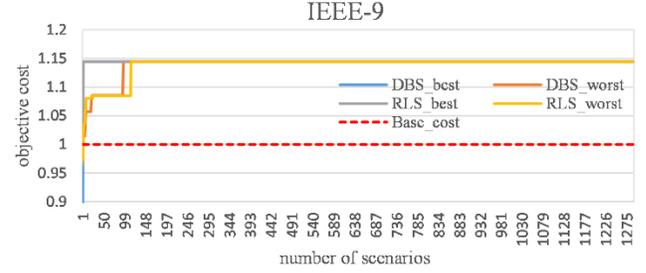

Fig.9 Performance Comparison of RLS and DBS in IEEE-9 System

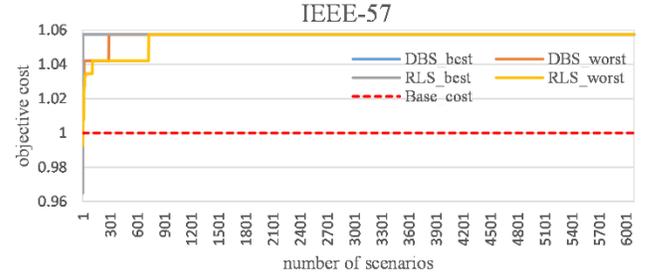

Fig.10 Performance Comparison of RLS and DBS in IEEE-57 System

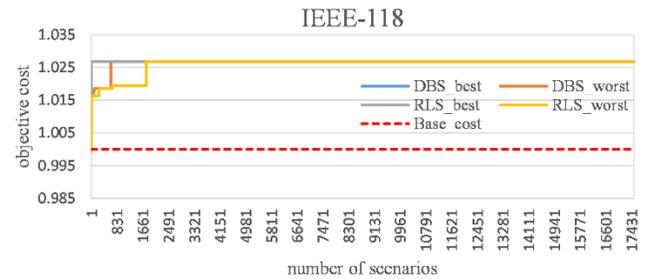

Fig.11 Performance Comparison of RLS and DBS in IEEE-118 System

For the figures above, the *x*-axis denotes the number of scenarios required at each round of solving CC-PSO-ES, and the *y*-axis is the objective cost of CC-PSO-ES. At each round of computation in verification process, a new scenario is added sequentially. In other words, each figure depicts the relationship between the objective cost and number of scenarios, i.e., how the objective cost changes as the number of scenarios are added sequentially. Both DBS and RLS have two lines, i.e., the best and worst cases, denoted as 'DBS_best' and 'DBS_worst', 'RLS_best' and 'RLS_worst', respectively. The dashed line represents the total cost without considering the uncertainty of REPL, denoted as 'Base_cost'. The objective cost is normalized by 'Base_cost'. The lines may become flat finally as the new scenarios are incorporated sequentially, which indicates that the optimal solution of CC-PSO-ES problem has been achieved.

## V. Conclusion and future work

This paper presents a novel solution method for solving the chance-constrained multi-period optimal power system operation (PSO) with battery energy storage (CC-PSO-ES), which is originally nonconvex and computationally intractable. The proposed method, which is based on the data-driven convex quadratic approximation (DDCQA) of ACPF and the strategic sampling, i.e., hybrid sampling methods, only uses a small number of scenarios without the pre-known PDF of the uncertainty of REPL. The DDCQA is modified through the generalized LASSO and applied to address the nonconvex problem of ACPF constraints in CC-PSO-ES problem. Unlike the RS-based methods, the hybrid sampling methods (HSMs) are proposed with dissimilarity-based learning and reinforcement learning methods. HSMs determine a smaller sample size than the RS-based methods. Eventually, the originally intractable CC-PSO-ES is converted to a tractable convex quadratic optimization problem with few effective scenarios. In our future work, we intend to test the proposed method in real-life large-scale power systems.

## VI. References

[1] Energy Storage Grand Challenge, 1st ed., The U.S. Department of Energy, Washington DC, 2020, pp. 12–20

[2] M. Asensio and J. Contreras, "Stochastic Unit Commitment in Isolated Systems With Renewable Penetration Under CVaR Assessment," *IEEE Trans. Smart Grid*, vol. 7, no. 3, pp. 1356-1367, May 2016, doi: 10.1109/TSG.2015.2469134.

[3] C. Zhao and Y. Guan, "Unified Stochastic and Robust Unit Commitment," *IEEE Trans. Power Syst.*, vol. 28, no. 3, pp. 3353-3361, Aug. 2013, doi: 10.1109/TPWRS.2013.2251916.

[4] F. Alismail, P. Xiong and C. Singh, "Optimal Wind Farm Allocation in Multi-Area Power Systems Using Distributionally Robust Optimization Approach," *IEEE Trans. Power Syst.*, vol. 33, no. 1, pp. 536-544, Jan. 2018, doi: 10.1109/TPWRS.2017.2695002.

[5] Zhi Chen, Daniel Kuhn, Wolfram Wiesemann, "Data-Driven Chance Constrained Programs over Wasserstein Balls", 2018, arXiv:1809.00210.

[6] H. Zhang and P. Li, "Chance Constrained Programming for Optimal Power Flow Under Uncertainty," *IEEE Trans. Power Syst.*, vol. 26, no. 4, pp. 2417-2424, Nov. 2011, doi: 10.1109/TPWRS.2011.2154367.

[7] A. Pena-Ordieres, D. K. Molzahn, L. Roald and A. Waechter, "DC Optimal Power Flow with Joint Chance Constraints," *IEEE Trans. Power Syst.*, doi: 10.1109/TPWRS.2020.3004023.

[8] Q. Li, "General Scenario Program: Application in Smart Grid Optimization under Endogenous Uncertainty." *IEEE Trans. Power Syst*. Under review. arXiv:2104.13494 (2021).

[9] M.C. Campi, S. Garatti, "The Exact Feasibility of Randomized Solutions of Uncertain Convex Programs", *SIAM J. Optim.,* vol. 19, no. 3, pp:1211–1230, 2008. https://doi.org/10.1137/07069821X.

[10] Algo Carè, Simone Garatti, Marco C. Campi, "FAST—Fast Algorithm for the Scenario Technique", *Operations Research*, vol. 62, no. 3, pp. 662–671, June 2014.

[11] John Eason, Selen Cremaschi, "Adaptive sequential sampling for surrogate model generation with artificial neural networks", Computers & Chemical Engineering, vol. 68, pp: 220-232, September 2014.

[12] Chen, Xinjia. "Exact computation of minimum sample size for estimation of binomial parameters," *Journal of Statistical Planning and Inference*, vol. 141, no.8, pp: 2622-2632, 2011.

[13] T. Alamo, R. Tempo, and E. F. Camacho, "Randomized Strategies for Probabilistic Solutions of Uncertain Feasibility and Optimization Problems," *IEEE Trans. Autom. Control*, vol. 54, no. 11, pp. 2545-2559, Nov. 2009, doi: 10.1109/TAC.2009.2031207.

[14] R. Hu, Q. F. Li, "Efficient Solution Strategy for Chance Constrained Optimal Power Flow based on FAST and Data driven Convexification," arXiv:2105.05336, 2021. https://arxiv.org/abs/2105.05336.

[15] Pless, Robert, and Richard Souvenir. "A survey of manifold learning for images." *IPSJ Transactions on Computer Vision and Applications*, vol. 1, pp: 83-94, 2009.

[16] Xiang, Zhengliang, Yuequan Bao, Zhiyi Tang, and Hui Li. "Deep reinforcement learning-based sampling method for structural reliability assessment." *Reliability Engineering & System Safety*, vol. 199, pp:106901, 2021.

[17] S. M. Fatemi, S. Abedi, G. B. Gharehpetian, et al., "Introducing a Novel DC Power Flow Method with Reactive Power Considerations." *IEEE Trans. Power Syst.*, vol. 30, no. 6, pp. 3012- 3023, Nov. 2015.

[18] R. Hu, Q. Li and S. Lei, "Ensemble Learning based Linear Power Flow," *2020 IEEE Power & Energy Society General Meeting (PESGM)*, Montreal, QC, 2020, pp. 1-5, doi: 10.1109/PESGM41954.2020.9281793.

[19] R. Jabr, "Radial distribution load flow using conic programming." *IEEE Trans. Power Syst.*, vol. 21, no. 3, pp. 1458–1459, Aug. 2006.

[20] X. Bai, H. Wei, K. Fujisawa, and Y. Wang, "Semidefinite programming for optimal power flow problems." *Int. J. Elect. Power Energy Syst.*, vol. 30, no. 67, pp. 383–392, 2008.

[21] Q. Li , R. Ayyanar, and V. Vittal. "Convex optimization for DES planning and operation in radial distribution systems with high penetration of photovoltaic resources." *IEEE Transactions on Sustainable Energy* 7, no. 3 (2016): 985-995.

[22] C. Coffrin, H. Hijazi, et al., "The QC Relaxation: A Theoretical and Computational Study on Optimal Power Flow." *IEEE Trans. Power Syst.,* vol. 31, no. 4, pp. 3008–3018, July. 2016.

[23] D. Molzahn and I. Hiskens, "Sparsity-exploiting moment-based relaxations of the optimal power flow problem." *IEEE Trans. Power Syst.*, vol. 30, no. 6, pp. 3168–3180, Nov. 2015.

[24] Q. Li, and V. Vittal. "Convex hull of the quadratic branch AC power flow equations and its application in radial distribution networks." *IEEE Trans. Power Syst.*, vol.33, no. 1, pp.839-850, Jan. 2017.

[25] Q. Li, "Uncertainty-Aware Three-Phase Optimal Power Flow Based on Data-Driven Convexification." *IEEE Transactions on Power Systems* 36, no. 2 (2021): 1645-1648.

[26] R. Hu, Q. Li and F. Qiu, "Ensemble Learning Based Convex Approximation of Three-Phase Power Flow," *IEEE Trans. Power Syst.*, doi: 10.1109/TPWRS.2021.3055481.

[27] B. Kocuk, S. S. Dey, X. A. Sun, "Inexactness of SDP Relaxation and Valid Inequalities for Optimal Power Flow." IEEE Trans. Power Syst., vol. 31, no. 1, pp. 642-651, Jan. 2016.

[28] Z.Y. Wang, D.S. Kirschen, B.S. Zhang, "Accurate Semidefinite Programming Models for Optimal Power Flow in Distribution Systems." arXiv preprint arXiv:1711.07853, 2017.

[29] M. Gan, G. Chen, L. Chen, and C. L. P. Chen, "Term Selection for a Class of Separable Nonlinear Models," in IEEE Transactions on Neural Networks and Learning Systems, vol. 31, no. 2, pp. 445-451, Feb. 2020, doi: 10.1109/TNNLS.2019.2904952.

[30] V. Roth, "The generalized LASSO," in IEEE Transactions on Neural Networks, vol. 15, no. 1, pp. 16-28, Jan. 2004, doi: 10.1109/TNN.2003.809398.

[31] Daniel Bienstock, Michael Chertkov, and Sean Harnett, "Chance-Constrained Optimal Power Flow: Risk-Aware Network Control under Uncertainty," *SIAM Rev.*, vol.56(3), pp:461–495, 2014.

[32] Aien, Morteza, Ali Hajebrahimi, and Mahmud Fotuhi-Firuzabad. "A comprehensive review on uncertainty modeling techniques in power system studies." Renewable and Sustainable Energy Reviews, vol. 57, no.5, pp: 1077-1089, May 2016.

[33] Yang, Mao, Chaoyu Shi, and Huiyu Liu. "Day-ahead wind power forecasting based on the clustering of equivalent power curves." *Energy*, vol.218, no.5, pp: 119515, 2021.